\documentclass[aps,pre,reprint,superscriptaddress]{revtex4-2}
\usepackage{graphicx}
\usepackage{amsmath, cancel, color, soulutf8, float}

\begin{document}

\title{Continuous to intermittent flows in growing granular heaps}

\author{L. Alonso-Llanes}
\thanks{These two authors contributed equally}
\affiliation{Group of Complex Systems and Statistical Physics, Physics Faculty, University of Havana, 10400 Havana, Cuba}
\affiliation{Université de Strasbourg, CNRS, Institut Terre et Environnement de Strasbourg, UMR7063, 67000 Strasbourg, France}
\author{E. Mart\'{\i}nez}
\thanks{These two authors contributed equally}
\affiliation{Department of Physics, NTNU, NO-7491 Trondheim, Norway}
\author{A. J. Batista-Leyva}
\affiliation{Group of Complex Systems and Statistical Physics, Physics Faculty, University of Havana, 10400 Havana, Cuba}
\affiliation{Instituto Superior de Tecnologías y Ciencias Aplicadas (InSTEC), University of Havana, 10400 Havana, Cuba}
\author{R. Toussaint}
\email[]{renaud.toussaint@unistra.fr}
\affiliation{Université de Strasbourg, CNRS, Institut Terre et Environnement de Strasbourg, UMR7063, 67000 Strasbourg, France}
\affiliation{SFF PoreLab, The Njord Centre, Department of Physics, University of Oslo, P.O. Box 1074 Blindern, 0316 Oslo, Norway}
\author{E. Altshuler}
\email[]{ealtshuler@fisica.uh.cu}
\affiliation{Group of Complex Systems and Statistical Physics, Physics Faculty, University of Havana, 10400 Havana, Cuba}

\date{}

\begin{abstract}
If a granular material is poured from above on a horizontal surface between two parallel, vertical plates, a sand heap grows in time. For small piles, the grains flow smoothly downhill, but after a critical pile size $X_c$, the flow becomes intermittent: sudden avalanches slide downhill from the apex to the base, followed by an ``uphill front" that slowly climbs up, until a new downhill avalanche interrupts the process. By means of experiments controlling the distance between the apex of the sandpile and the container feeding it from above, we show that $X_c$ grows linearly with the input flux, but scales as the square root of the feeding height. We explain these facts based on a phenomenological model based on the experimental observation that the flowing granular phase forms a ``wedge" on top of the static one, differently from the case of stationary heaps. Moreover, we demonstrate that our controlled experiments allow to predict the value of $X_c$ for the common situation in which the feeding height decreases as the pile increases in size.
\end{abstract}

\maketitle

\section{Introduction}

Granular matter is known to exhibit many unexpected behaviors \cite{Knight1993PhysRevLett, Melo1994PhysRevLett, LePennec1996PhysRevE, Conway2003PhysRevLett, Taberlet2004EurophysLett, Elbelrhiti2005Nature, Martinez2007PhysRevE, Johnsen2008PhysRevE, Pacheco2011PhysRevLett, Niebling2012PhysRevE, Altshuler2014GeophysResLett, Eriksen2018PhysRevE, Turquet2019GeophysResLett, DiazMelian2020PhysRevLett, Espinosa2022GranMatt} since, depending on the way it is handled, it behaves as a solid, a liquid or a gas \cite{Jaeger1996RevModPhys,Andreotti2013Book}. Most granular flows encountered in nature, and industry, are located in the liquid regime, where the material is dense but still flows as a fluid. This regime have received no little attention from the scientific community in the search for constitutive laws capable of reproducing the diversity of observed behaviors \cite{Douady1999EuroPhysJournalB,Khakhar2001JFluidMech,MiDi2004EuroPhysJournalE, Jop2005JFluidMech, Jop2006Nature,Forterre2008AnnuRevFluidMech,Jop2015ComptesRendusPhys}. However, one behavior in particular has been poorly studied, and it constitutes the focus of our paper: the transition from continuous to intermittent flows (CIT).

The phenomenon was first described in the so-called rotating drum geometry, i.e., for a cylinder partially filled with granular matter rotating around its axis of symmetry, which lies in a plane perpendicular to the force of gravity. If this device rotates slowly enough, the particle flow near the free surface is intermittent producing avalanches. However, for higher rotational velocities, the particles establish a continuous flow such that the free surface shows a stationary profile, as reported by Rajchenbach \cite{Rajchenbach1990PhysRevLett}.

Shortly after, Lemieux and Durian showed the existence of a continuous to intermittent transition in the flow of glass beads down a stationary heap, confined between plane-parallel walls, as the input flux decreased \cite{Lemieux2000PhysRevLett}, which was corroborated by Jop \textit{et al.} when studying the role of confinement in this type of geometry \cite{Jop2005JFluidMech}.

\begin{figure}[H]
\centering
\includegraphics[scale=0.38]{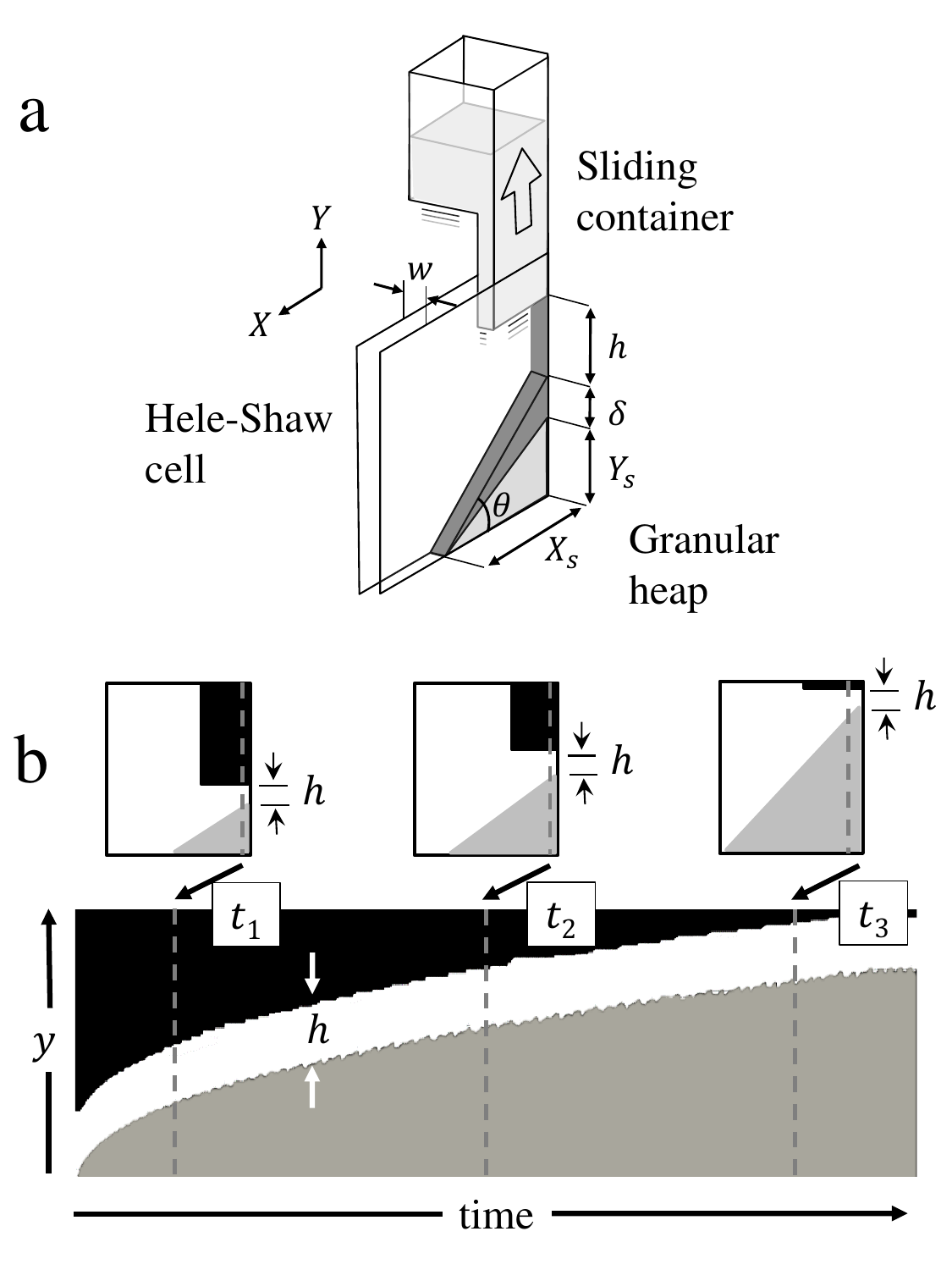}
\caption{Experimental setup. (a) Sketch of the experimental setup, where the sliding container allows to control the value of the grain feeding height, $h$ (the darker color indicate the mobile grains) (b) Diagram showing three stages of the heap growth at constant $h$, corresponding to times $t_1$, $t_2$ and $t_3$. The upper side shows a lateral view of the Hele-Shaw cell including a vertical line of pixels that allows to construct the spatial-temporal diagram shown in the bottom panel. The diagram, which corresponds to a real experiment, illustrates the constancy of $h$.}
\label{fig:Fig1}
\end{figure}

Altshuler and co-workers \cite{Altshuler2003PhysRevLett} also observed a CIT in narrow sand rivers moving through conical piles in various configurations: by increasing the pile base size, grown with constant input flux, or by varying the input flux in both open and closed piles with a constant size base. In another study by this group, they observed that the transition was maintained after reducing the pile dimension by confining it in a Hele-Shaw cell \cite{Altshuler2008PhysRevE}.

More recently, Fischer \textit{et al.} \cite{Fischer2009PhysRevLett} found that for a given range of rotational speeds, rotating drums exhibit a progressive transition of flow by temporal intermittency, where spontaneous and erratic changes from one regime to another occur.

Here we study the CIT in a growing granular heap --in which the control parameter is its size-- where the drop height of the grains is controlled.Our experiments reveal that the horizontal size of the pile where the CIT takes place, $X_c$, depends linearly on the input flux and varies as the squared root of the dropping height. We explain those scalings in terms of a model based on previous knowledge on dense granular flows. On the other hand, we show that it is possible to predict $X_c$ without controlling the dropping height, based on the data obtained by controlling it.

\section{Experimental}

A sand heap is grown into a Hele-Shaw cell of width $w=12$ mm consisting in two vertical glass plates perpendicular to a flat horizontal surface of $28$ cm long. One of the thin vertical sides of the cell is closed by a glass wall, and the opposite one is open. A thin stream of sand is dropped through a modifiable rectangular slit made in the bottom of a parallelepiped-shaped container held on top of the cell, so the sand enters parallel and near the thin, closed vertical wall of the Hele-Shaw cell. As a result, a sand heap forms that grows from the closed vertical wall to the open wall, as depicted in Fig. \ref{fig:Fig1}(a). The input flux $F_{in}$ is obtained by processing images taken with a camera at a capture rate of $30$ fps and a resolution of $1024 \times 768$ pixels: the total pile area is computed in each frame and $F_{in}$ is determined as the slope of its dependence with time. $F_{in}$ is defined as $Q_{in}/w$, where $Q_{in}$ is the volumetric input flux. We used sand with a high content of silicon oxide and an average grain size of $d = 100$ $\mu$m from Santa Teresa (Pinar del Río, Cuba) \cite{Altshuler2003PhysRevLett, Martinez2007PhysRevE, Altshuler2008PhysRevE}.

A feature that distinguishes this setup from most ones reported in the literature is the fact that the distance between the point of delivery of the sand and the upper part of the heap, $h$, can be held constant in time (within a range from $1$ cm to $15$ cm) due to the vertical movement of the sand container, see Fig. \ref{fig:Fig1} (b). This is particularly important for growing heaps since this distance changes during the experiments if the point of delivery is fixed, which is not the case in the commonly studied stationary heaps. The control of the deposition height is achieved thanks to a feedback system: when a laser light is interrupted by the growing tip of the heap, a signal is sent to a motor that rises the position of the sand container until a new laser beam detection \cite{Dominguez2015RevCubFis, Alonso2017RevCubFis}. The constancy of $h$ is illustrated in the spatial-temporal diagram shown in Fig. \ref{fig:Fig1}(b) taken along a vertical line near the taller side of the heap. If the container is kept at a fixed height relative to the laboratory, the heap grows in a ``conventional way", i.e., with $h$ decreasing as the height of the pile increases, which may be more interesting for engineering applications.

\section{Results and discussion}
\subsection{Presentation of the continuous to intermittent transition}

\begin{figure}
\includegraphics[scale=0.30]{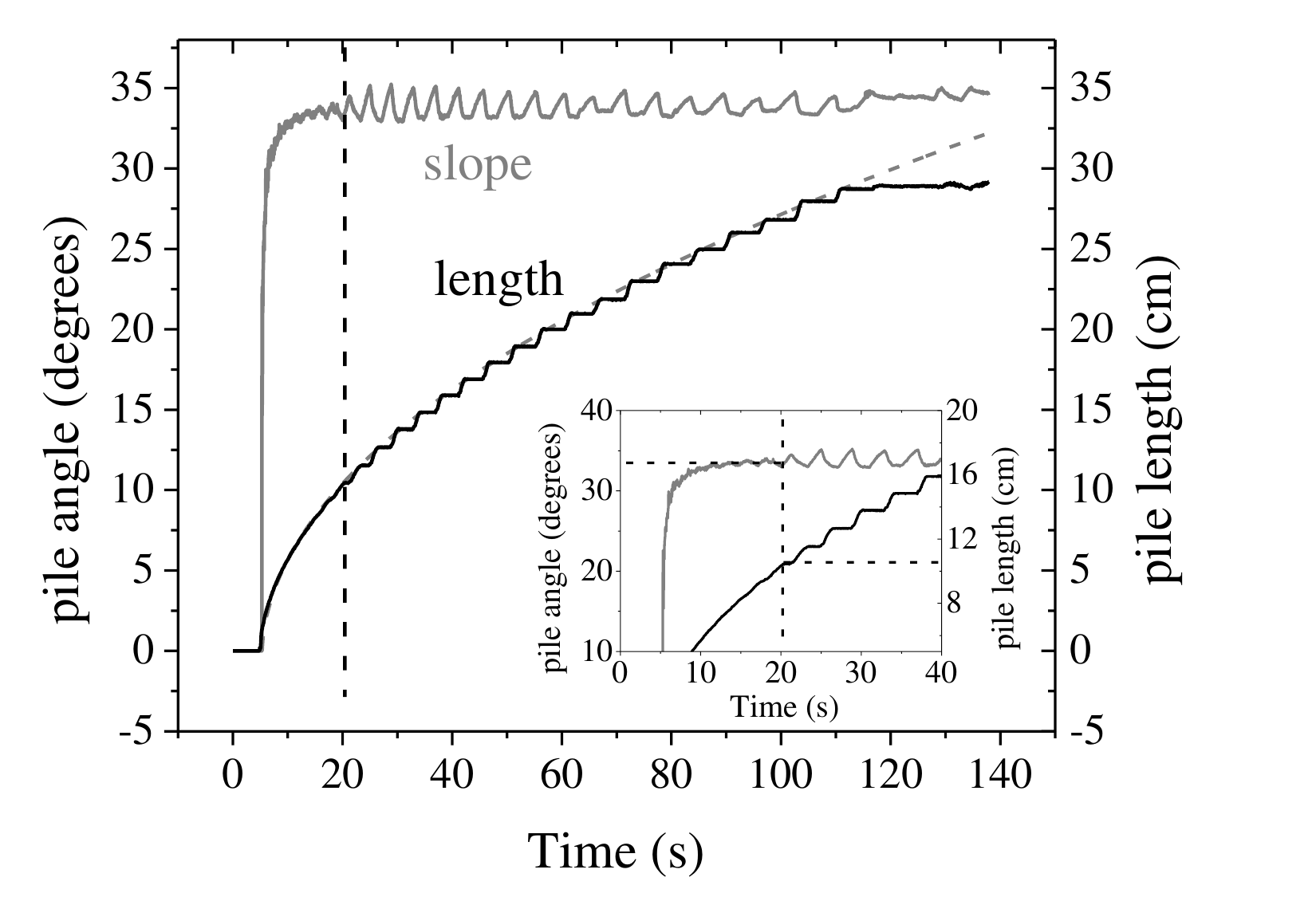}
\caption{Quantifying the continuous to intermittent flow transition. The gray continuous line corresponds to the temporal evolution of the average slope of the pile, while the black continuous line represents its horizontal length very near the bottom of the Hele-Shaw cell. The light gray line (dashed) is a fit following the law $x \sim t^{1/2}$. The transition occurs at $X_c \approx 10.3$ cm, $t_c \approx 20$ s and with a repose angle of $\theta_s \approx 33.5 ^{\circ}$, as clearly seen in the zoom shown in the inset. The data represented in the graphs correspond to a fixed input flux of $F_{in}=2.68$ cm$^2$/s and a constant deposition height of $h\approx1.5$ cm.}
\label{fig:Fig2}
\end{figure}

As the piles grow, a transition between continuous to intermittent behavior is observed. From the time the first grains hit the floor of the Hele-Shaw cell, up to a certain horizontal length $X_c$ of the bottom of the heap, the flow of grains is continuous: grains flow down the heap forming a fluid layer next to the free surface of the pile, which is approximately straight. As the horizontal length of the pile grows above $X_c$, an intermittent regime is reached: an ``avalanche" of grains slides down the hill from the upper side of the pile all the way to the lower side, and stops, forming a step-front that moves uphill as new grains are fed into the pile. Once the uphill step-front is within a few inches of the top of the pile, a new downhill avalanche takes place, and so on. The continuous to intermittent flow transition occurring at $X=X_c$ was originally reported for experiments in which $h$ was not controlled \cite{Altshuler2008PhysRevE}, and has been recently seen by us for controlled height experiments \cite{Alonso2017RevCubFis}.

For all combinations of deposition heights and fluxes studied, the transition can be visualized as in Fig. \ref{fig:Fig2}, which is based on a measurement using a fixed input flux of $2.68$ cm$^2$/s and a constant deposition height $h \approx 1.5$ cm. The first quantity plotted is the length of the base along the horizontal axis near the bottom of the Hele-Shaw cell \footnote{The distance from the bottom is taken as half the width of the input jet of sand feeding the pile} as time goes by. We have also plotted the temporal evolution of the average angle of the free surface relative to the horizontal. Both curves complement each other in order to extract the transition coordinates, as the vertical dashed line suggests.

The horizontal length of the pile grows as mass conservation dictates and stops for the first time after reaching a value of about 10.3 cm, from which it continues as steps. On the other hand, the average slope of the pile describes a similar behavior: it grows in time and then starts to oscillate around an average angle of approximately $33.5$ degrees. Note that shortly before the onset of the oscillations the average slope has some noisy oscillations with relative small amplitude that may be associated with a transient in which spontaneous transitions occur between both regimes as observed in rotating drums \cite{Fischer2009PhysRevLett}.
Both changes in behavior occur about 20 s after the beginning of the experiment (see vertical dashed line). That is just the moment when the growth of the horizontal length of the heap goes from a continuum to an intermittent regime: the horizontal steps correspond to the time intervals during which a step-front climbs uphill after each downhill avalanche, keeping constant the length of the pile. This feature is also seen in the curve of the angle where an increase in its value represents the climb of the step-front while a decrease represents the occurrence of one avalanche. Notice that after $t \approx 110$ s the pile bottom reaches the outlet, turning into a system widely studied in the literature --a stationary heap-- which we do not discuss here.

\subsection{Dependence of the transition coordinates on the input flux and 
deposition height}

\begin{figure}
\includegraphics[scale=0.5]{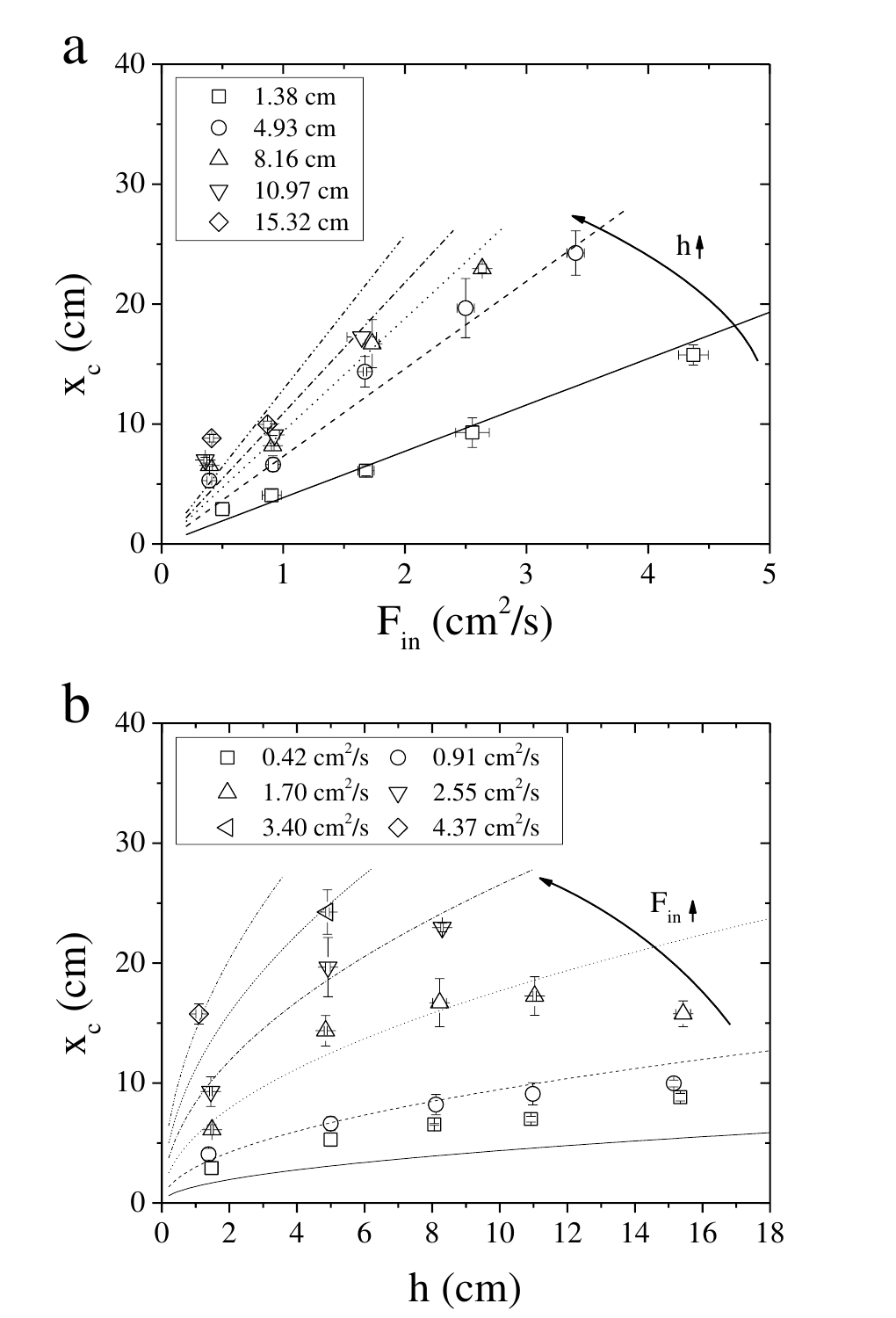}
\caption{Dependencies of the transition length on the input flux and deposition height. Transition length as a function of (a) the input flux, for different deposition heights and (b) the deposition height, for different input fluxes. Symbols represent experimental data while the lines the theoretical expression proposed in (\ref{eq:transition}). Note that the bottom of the cell has a finite size of 28 cm, thus for certain parameter values the base of the pile reaches the open edge of the cell before transition occurs. This leads to some curves shown having fewer data points than others.}
\label{fig:Fig3}
\end{figure}

Fig. \ref{fig:Fig3} shows the dependence of the pile transition lengths ($X_c$) on the input flux ($F_{in}$) and the deposition heights ($h$). Fig. \ref{fig:Fig3}(a) indicates that the dependence on the transition length is linear with the input flux. On the other hand, Fig. \ref{fig:Fig3}(b) shows that the transition length scales as the square root of the deposition height. These dependencies can be understood by means of a simple model of the growing heap explained in detail in the Appendix, see Fig. {\ref{fig:sketch}}. Fig. {\ref{fig:sketch}}(a) shows an image that reveals the geometry of the flowing layer during the continuous regime. The experiment was performed on silica sand with average diameter 1.5 mm, and a video was taken with a fast camera at 1000 fps. Note that, differently from the case of stationary heaps, the flowing layer is wedge-shaped in growing piles, where the wedge point is at the lower end of the pile. Fig. {\ref{fig:sketch}}(b) shows a schematic based on the observed geometry, showing all the parameters used in our model.

Let us consider a flow of incompressible grains ($\rho = $ constant) on the free surface of a growing heap during the continuous regime. Inside a control volume that moves with velocity {$v_{\perp}$} to include only the fluid layer, the equations of mass and linear momentum (in the $x$-direction) conservation of the flowing grains are as follows:

\begin{widetext}
\begin{equation} \label{eq:mass_momentumx}
\begin{split}
    \rho \frac{\partial \delta}{\partial t} + v_{\parallel} \frac{\partial \delta}{\partial x} &= - v_{\perp} \\ 
    \rho \frac{\partial \left( v_{\parallel} \delta  \right)}{\partial t} + \rho v_{\parallel}^2 \frac{\partial \delta }{\partial x} &= - \frac{\partial \left(\delta  \tau_{xx}\right)}{\partial x} + \tau_{yx} + \tau_{zx}\frac{\delta }{w} + \rho g \delta  \sin\theta + (\rho v_x v_{\perp})|_{y=0}
\end{split}
\end{equation}
\end{widetext}  

In these equations $\delta$ is the thickness of the flowing layer of grains, and $w$ its width,$v_{\parallel} = \frac{1}{\delta w}\int_{0}^{w}\int_{0}^{\delta} v_x dzdy$ is the averaged velocity down the heap, where $v_x$ is considered constant along the fluid layer as it is suggested by experimental measurements \cite{Lemieux2000PhysRevLett, Jop2005JFluidMech}. $\tau_{xx}$ is the normal stress while $\tau_{yx} $ and $\tau_{zx}$ are the shear stresses at the interface between the flowing and static layers, and at the walls, respectively.

We make some other assumptions in order to simplify equations in ({\ref{eq:mass_momentumx}}). The normal stress $\tau_{xx}$ is considered as the dynamic pressure, equal to $\frac{1}{2} \rho v_{\parallel}^2 $, and its variation in the flow direction ($x$) is neglected since changes in the layer thickness are small. The shear stresses $\tau_{yx} = \rho g \delta \cos\theta\tan\theta_s$ and $\tau_{zx} = \mu_w \rho g \delta \cos\theta$ are assumed as Coulomb's frictional stresses where $\tan\theta_s$ is the static friction of the pile $\mu_s$ and $\mu_w$ the dynamic friction coefficient of the grains with the walls. $\tau_{zx}$ is taken as suggested in {\cite{Jop2005JFluidMech}}. $v_x|_{y=0}$ is expected to be equal or close to $0$, so, $\rho v_x v_{\perp}$ is neglected. The quasi-steady solution of the obtained equations, the one in which a quasi-steady flow is considered where the time variations of $\delta$ and $v_{\parallel}$ are approximately equal to 0, gives (See Appendix {\ref{Appendix.A}} for details),

\begin{equation}
L_x = \frac{3 Q_{in} \alpha \sqrt{2g h}}{2 g \bar\delta \cos\theta \left(-\tan\theta + \mu_w \frac{\bar\delta }{w} + \tan\theta_s\right) w}
\label{eq:mass_momentumx3}
\end{equation}

Here $\bar\delta = (1/L_x)\int_0^{L_x} \delta dx$ and $Q_{\perp} = Q_{in} = v_{\perp} L_{x} w $, where $L_{x} = \frac{X}{\cos\theta}$ is the length of the interface between the flowing and static layers, and $v_{\parallel} = \alpha v_{in} = \alpha \sqrt{2g h} \sin\theta$, being $v_{in}$ the velocity of the incoming grains (free fall) and $\alpha < 1$ a dimensionless constant that accounts for the energy loss after the impact on the tip of the pile of the grains coming from the container.

At the transition ($\theta \to \theta_s$) the thickness of the fluid layer reaches a minimum $\bar\delta=\delta_{stop}$, after which the flowing grains cannot ``continually cover" the whole surface of the pile, therefore intermittency starts \cite{Lemieux2000PhysRevLett,Jop2005JFluidMech,MiDi2004EuroPhysJournalE}. At this point, Eq. (\ref{eq:mass_momentumx3}) transforms into the following:

\begin{equation}
    X_c = \frac{3 \alpha w \sin\theta_s}{ \sqrt{2g} \mu_w \delta _{stop}^2}F_{in}h^{1/2}
    \label{eq:transition}
\end{equation}

\begin{figure}
\includegraphics[scale=0.45]{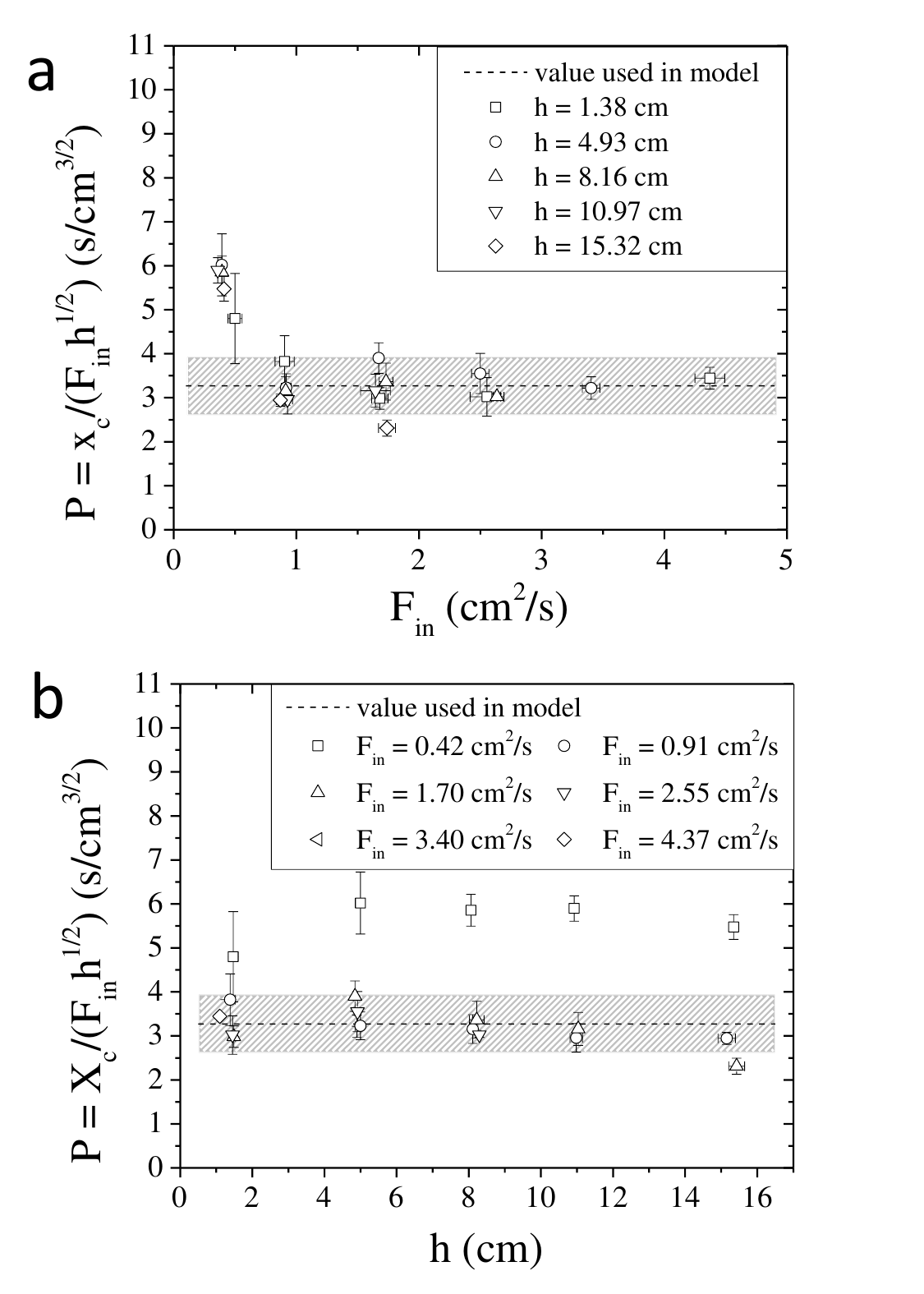}
\caption{Quantitative comparison between experiments and model. Dependence with the input flux (a) and with the deposition height (b) of the proportionality constant in Eq. (\ref{eq:transition}), $P$. Black dashed lines represent the value of $P = X_c/(F_{in}\sqrt{h})$ used in our model to reproduce the experimental data and the height of the gray rectangles represents its standard error. $P_{model} = 3.27 \pm 0.65$ s/cm$^{3/2}$.}
\label{fig:Fig4}
\end{figure}

Fig. \ref{fig:Fig4} shows indeed that $P=X_c/(F_{in}\sqrt{h})$, which is the prefactor of Eq. ({\ref{eq:transition}}), is roughly constant with no systematic dependence on the control parameters $F_{in}$ and $h$. There is only a weak dependence observed for the lowest flux $F_{in}=0.42$ cm$^2$/s for which this quantity slightly rises with $h$ up to a value around $6$ s/cm$^{3/2}$. Thus, Eq. ({\ref{eq:transition}}) explains the general shape of the dependencies displayed in Fig. \ref{fig:Fig3}(a) and (b). These laws can be reproduced quantitatively using values shown in Table {\ref{table:quantities}}, that correspond to a prefactor value of $P=X_c/(F_{in}\sqrt{h})= 3.27 \pm 0.65$ s/cm$^{3/2}$. $\theta_s$ is taken from the experimental data while $\mu_w$ and $\delta_{stop}$ are taken from the literature typical values of these quantities. That gives a value of $\alpha = 0.30$ which is our free parameter and suggests that the 70 $\%$ percent of the potential energy of the grains is dissipated due to the impact at the tip of the pile. 

\begin{table}[ht]
\centering 
\begin{tabular}{c c} 
    \hline 
    Magnitude & Value \\ [0.5ex] 
    \hline 
    $\theta_s$ & 33.5\textdegree \\ 
    $\mu_w$ & 0.18 \cite{Jop2005JFluidMech}\\
    $\delta _{stop}$ & 15 $d$ = 0.15 cm \cite{Lemieux2000PhysRevLett} \\
    $\alpha$ & 0.30 \\ [1ex] 
    \hline 
    \end{tabular}
    \caption{Values of the quantities in eq. (\ref{eq:transition})} 
    \label{table:quantities} 
\end{table}

\begin{figure}[ht]
    \includegraphics[scale=0.5]{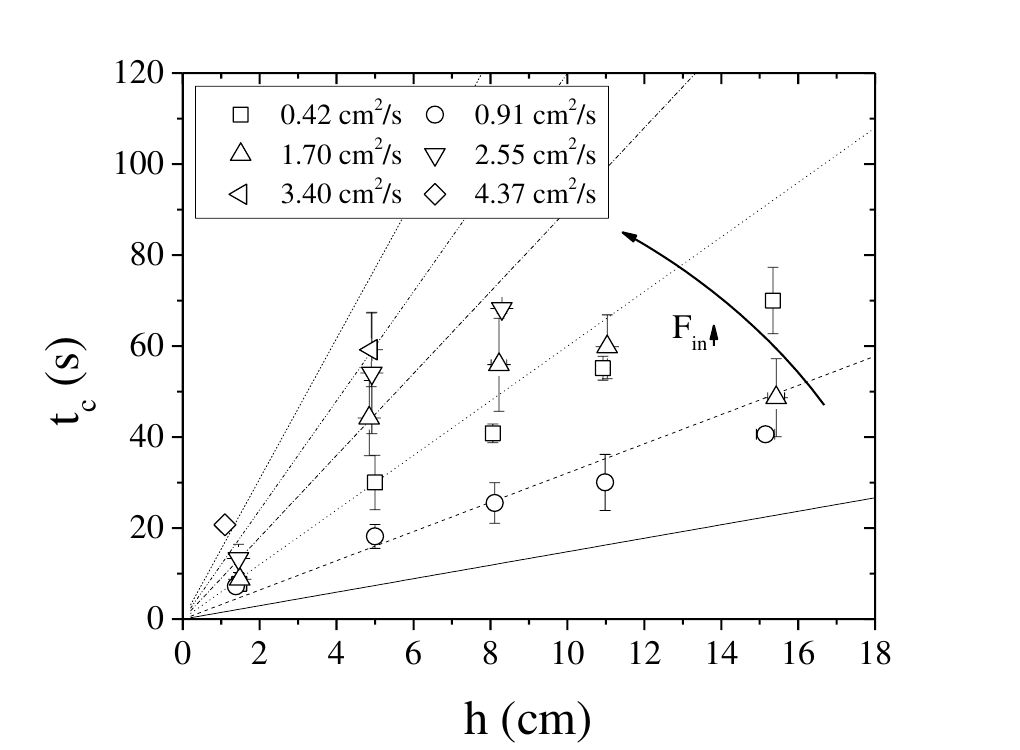}
    \caption{Transition time dependency on the deposition 
    height. Transition time as a function of deposition height for 
    different input fluxes.} 
    \label{fig:Fig3.1}
\end{figure}

The agreement between the theoretical expression and the experimental measurements of $X_c$ is satisfactory for all values of $F_{in}$ and $h$ tested, as seen on Fig. \ref{fig:Fig3}(a) and \ref{fig:Fig3}(b), comparing the theoretical expressions (lines) and experimental results (symbols). The only significant discrepancy is for the lowest input flux, where the model underestimates the value of $X_c$ by a factor between 1.5 and 1.8. Apart from this outlying results at the lowest flux, and a point at $F_{in}=1.7$ cm$^2$/s and $h=15$ cm, all other experimental values of $X_c$ lie within 18\% of the theoretical value with fixed $\theta_s$, $\mu_w$, $\alpha$ and $\delta_{stop}$. 

Furthermore, taking into account that the conservation of mass imposes that the horizontal length of the pile scales with time as $t^{1/2}$ (see the fit displayed in Fig. \ref{fig:Fig2}), Eq. (\ref{eq:transition}) immediately explains (at least semi-quantitatively) the experimental behavior illustrated in Fig. \ref{fig:Fig3.1}. It also predicts the behavior of $t_c$ as function of the input flux which, following the model, is given by (see Appendix \ref{Appendix.A}),

\begin{equation}
t_c = \frac{9 \alpha^2 w^2 \sin^2\theta_s \tan\theta_s}{4g \mu^2_w \delta_{stop}^4} F_{in} h
\label{eq:transition2}
\end{equation}

Here the agreement between experimental data and theory is satisfactory as well, although less than for the value of $X_c$. Note that, apart from the lowest flux and two points with $F_{in}=1.7$ cm$^2$/s, one with $h=5$ cm and another with $h=15$ cm, the experimental points lie within $23\%$ of the theoretical value. 

\subsection{Predicting the transition: from fixed-$h$ to variable-$h$ piles}

\begin{figure}[h]
\includegraphics[scale=0.45]{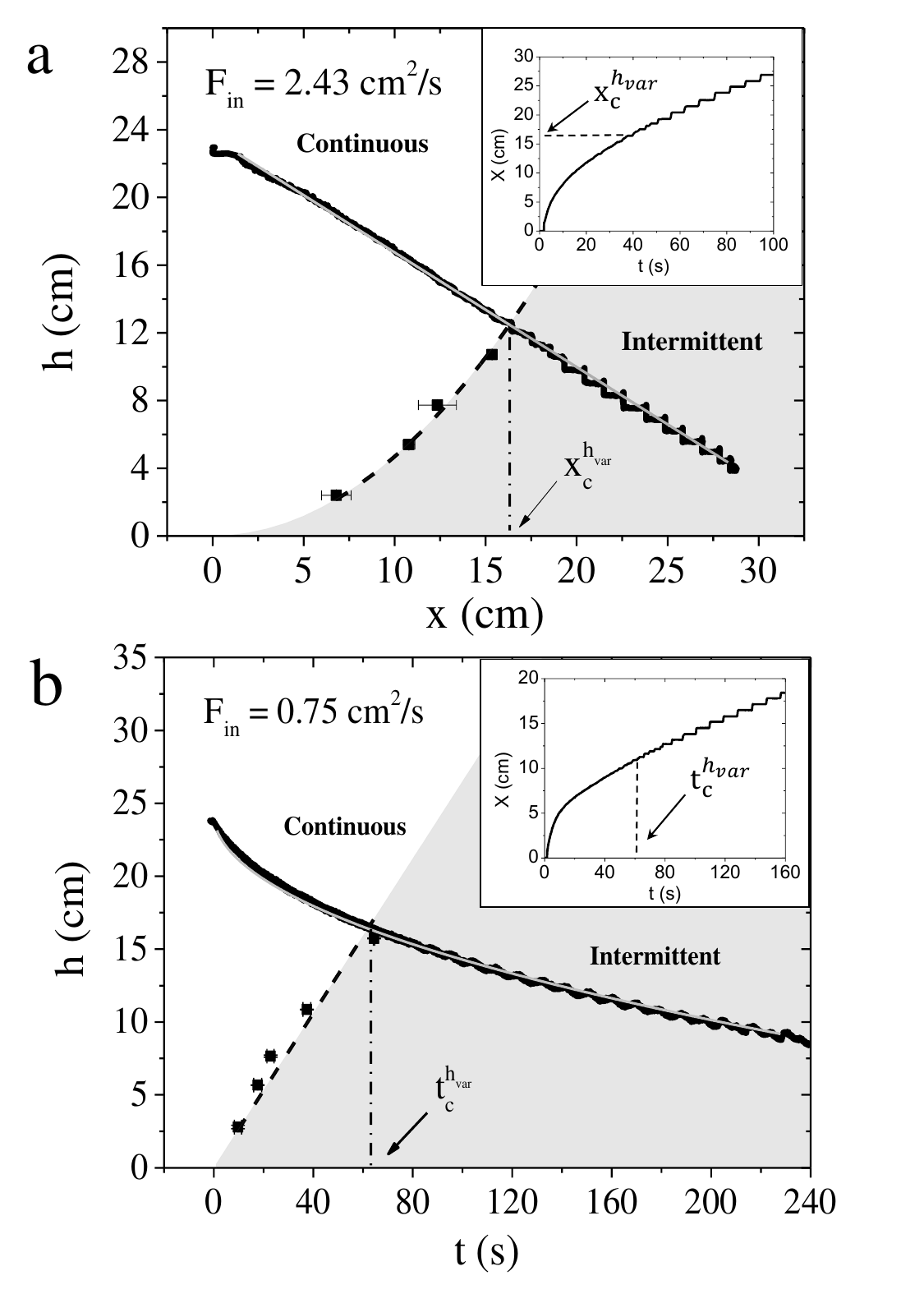}
\caption{Equivalence between fixed-$h$ and variable-$h$ experiments. The black solid lines show the deposition height for a variable-h experiment versus (a) the pile length and (b) the time, while the dashed increasing lines represent (a) a quadratic fit and (b) a linear fit to the dependence between the respective magnitudes at the transition for fixed-$h$ experiments (squares are experimental points). Gray solid lines follow the law (a) $h = 23.5$ cm $- 0.67 x$ and (b) $h = 23.5$ cm $- (0.99$ cm$\cdot$s$^{-1/2})\sqrt{t}$. The insets display the time evolution of the pile horizontal length for variable-$h$ experiments. (a) Input flow of $2.43$ cm$^2$/s (b) Input flow of $0.75$ cm$^2$/s.}
\label{fig:Fig5}
\end{figure}

Differently from most of previous research, our experiments show the CIT in a completely controlled way: not only is the input flux controlled, but also the deposition height is fixed. In typical situations where some granular material is poured to form a growing heap, it is poured from a container without changing its position, so $h$ decreases as the upper side of the heap grows. That behavior is showed for experiments we made with a fixed container and represented by the solid, decreasing lines in Fig. \ref{fig:Fig5}. 

Fig. \ref{fig:Fig5}(a) shows how the deposition height decreases as the length of the pile increases while Fig. \ref{fig:Fig5}(b) shows its temporal evolution. Both behaviors can be modeled by applying the mass conservation principle, and approximating the pile as a triangular shape with fixed angle $\theta_s$, height $Y$ and base $X$. Since $Y+h=h_0$, $Y=X\tan\theta_s$, and from mass conservation $XY=2Ft$, one obtains that $h$ varies with the pile length as $h_0 - X\tan\theta_s$ and varies in time as $h_0 - \sqrt{2t F \tan{\theta_s}}$, where $h_0$ is the initial height of the container over the bottom of the Hele-Shaw cell and $\theta_s$ the critical angle of the pile surface. These expressions correspond to the two gray lines shown in Fig. \ref{fig:Fig5} (a) and (b). Their agreement with the experimental data, the black continuous line, is noticeable.

Dashed lines in Fig. (\ref{fig:Fig5}) are constructed by fits to the experimental data where $X_c$ and $t_c$ were obtained in different experiments controlling $h$ (represented by squares). Each point of these lines corresponds to (a) $X_c^{h_{fix}}$ and (b) $t_c^{h_{fix}}$ for the different values of the fixed deposition height $h_{fix}$. The lines represent the interface between the continuous and the intermittent regimes for those input fluxes. Then, the length (time) at which the two curves intercept should match the transition length (time) for a free-$h$ experiment, i.e., $X_c^{h_{fix}}=X_c^{h_{var}}$ ($t_c^{h_{fix}}=t_c^{h_{var}}$). The insets, which correspond to piles grown without controlling the deposition height, show that our prediction is correct in each case. Furthermore, using that idea it can be determined at which fixed-$h$ the transition occurs at the same length or time for both kind of experiments, fixed-$h$ and variable-$h$. 
 
\section{Conclusions}

We have performed a systematic study of the surface flow on a granular heap where grains are added from a controlled height, $h$.
As the pile grows, the flow is first continuous, and then intermittent: for small piles, the free surface is smooth, but after reaching a horizontal pile size $X_c$ avalanches flow down the hill, and step-like fronts then climb uphill until a new downhill avalanche occurs. We have found that $X_c$ grows linearly with the input flux, $F_{in}$, and as $h^{1/2}$ with the deposition height $h$. We explain these facts based on a model where mass and momentum are conserved taking into account the basal friction of the flowing grains with the quasi-static part of the heap as well as the friction with walls. Importantly, the model assumes that the layer of flowing grains is wedge-shaped in growing heaps, a fact that we demonstrate experimentally. Moreover, by systematically comparing experiments with controlled $h$ and with non-controlled $h$ (as typically reported in the literature) we are able to predict the values of $X_c$ and $t_c$ in the latter case based on its values in the former.

\section{Acknowledgments}

We thank the support provided by Campus France Cuba, the French Embassy in Havana, the ITES, the University of Strasbourg, the INSU ALEAS program, the International Research Project France-Norway on Deformation Flow and fracture of disordered Materials IRP D-FFRACT and the Research Council of Norway through its Centres of Excellence funding scheme, project number 262644. Discussions with Reinaldo García are gratefully acknowledged.

\appendix
\section{Deriving equations (\ref{eq:transition}) and (\ref{eq:transition2})}\label{Appendix.A}

\begin{figure}[h]
    \centering
    \includegraphics[scale=0.5]{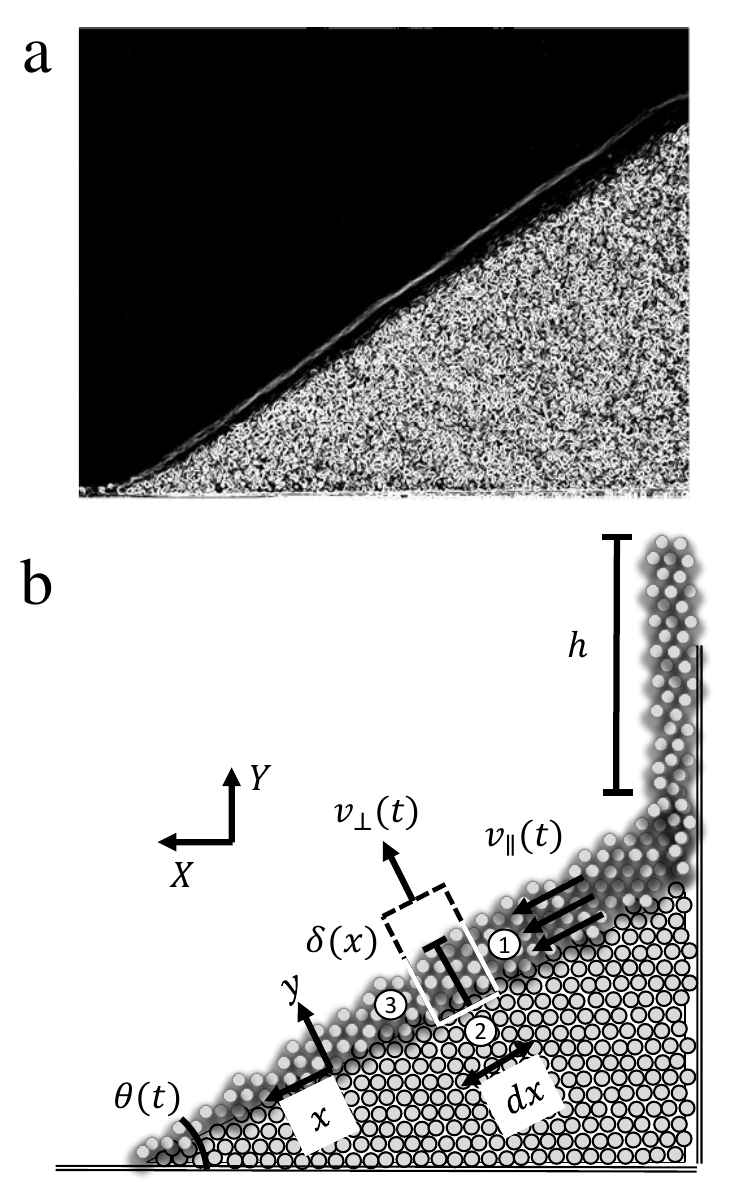}
    \caption{(a) Flowing layer in a typical experiment during the continuous regime. The image is obtained from the overlay of 100 images (corresponding to $1/10$ seconds) and then applying an edge detector. (b) Sketch used to derive the model equations. The region indicated with dashed lines represents the control volume (CV) while the solid white lines on top of them represent the control surfaces (CS) through which there is interchange of mass and momentum. The control volume moves with velocity $v_{\perp}$ to include only the fluid layer. The velocity $v_{\parallel} = \frac{1}{\delta w}\int_{0}^{w}\int_{0}^{\delta} v_x dzdy$.}
    \label{fig:sketch}
\end{figure}

If we consider a flow of incompressible grains ($\rho = $ constant) of thickness $\delta$ and width $w$ on the free surface of a growing heap, using the Reynolds' transport theorem we can write the depth-averaged mass conservation equation for a control volume (CV), as the one showed in Figure {\ref{fig:sketch}}. Here, $\vec{v}$ is measured relative to the control volume (CV) and the velocity along the $x$-direction considered constant $\frac{\partial v_{x}}{\partial x} = 0$ as it is suggested by experimental measurements \cite{Lemieux2000PhysRevLett,Jop2005JFluidMech}. 

\begin{widetext}
\begin{equation}
    \begin{split}
        \frac{dm}{dt} & = \frac{\partial}{\partial t} \left( \int_{CV} \rho \,dV \right) + \int_{CS} \rho (\vec{v} \cdot \vec{n}) \,dA \\
        0 & = \frac{\partial}{\partial t} \left(\rho w \delta dx \right) - \int \rho v_{\parallel} dA_1 + \int \rho v_{\perp} dA_2 + \int \rho v_{\parallel} dA_3 \\
        0 & = \rho \frac{\partial \delta}{\partial t} + \rho v_{\parallel} \frac{\partial \delta}{\partial x} + \rho v_{\perp}
    \end{split}
\end{equation}

The quasi-steady state solution gives 

\begin{equation} \label{eq:mass_final}
    v_{\parallel} \frac{\partial \delta}{\partial x} = - v_{\perp}
\end{equation}

Now, doing the same analysis but for the linear momentum conservation,

\begin{equation}
    \frac{d\vec{p}}{dt} = \frac{\partial}{\partial t} \left( \int_{CV} \vec{v} \rho \,dV \right) + \int_{CS} \vec{v} \rho (\vec{v} \cdot \vec{n}) dA 
\end{equation}

and writing its force balance equation for the $x$-direction, we get,

\begin{equation} \label{eq:force_balance}
\begin{split}
    \sum{F_x} & = \tau_{xx}\delta w - \left((\tau_{xx} + \frac{\partial \tau_{xx}}{\partial x}dx)(\delta +\frac{\partial \delta }{\partial x}dx)w\right) + \tau_{yx} w dx + 2\tau_{zx} \delta dx + \rho g \sin\theta \delta w dx\\ 
    & = w dx \left(- \frac{\partial \left(\delta \tau_{xx}\right)}{\partial x} + \tau_{yx} + \tau_{zx}\frac{\delta }{w} + \rho g \delta  \sin\theta \right) 
\end{split}
\end{equation}

In Eq. ({\ref{eq:force_balance}}) we take into account the normal and shear stresses in the x-direction on every control surface indicated in Fig. {\ref{fig:sketch}} as well as the shear stress on the walls confining the flow. Now, looking at the momentum exchange through the control surfaces,

\begin{equation} \label{eq:momentum_exchange}
\begin{split}
    \sum{F_x} & = \frac{\partial}{\partial t} \left( \int_{CV} \vec{v} \rho dV \right) + \int_{CS} \vec{v} \rho (\vec{v} \cdot \vec{n}) dA \\ 
    & = \frac{\partial \left( v_{\parallel} \rho w \delta  dx \right)}{\partial t} - \int v_{\parallel} \rho v_{\parallel} dA_1 + \int v_x \rho v_{\perp} dA_2 + \int v_{\parallel} \rho v_{\parallel} dA_3 \\
    & = w dx \left(\rho\frac{\partial \left(v_{\parallel} \delta  \right)}{\partial t} \right) - \rho v_{\parallel}^2 w \delta  - (\rho v_x v_{\perp})|_{y=0} w dx + \rho v_{\parallel}^2 w \left(\delta  + \frac{\partial \delta }{\partial x}dx\right)  \\
    & = w dx \left(\rho\frac{\partial \left( v_{\parallel} \delta  \right)}{\partial t} - (\rho v_x v_{\perp})|_{y=0} + \rho v_{\parallel}^2 \frac{\partial \delta }{\partial x}\right)
\end{split}
\end{equation}

Using Eq. \ref{eq:force_balance} and \ref{eq:momentum_exchange},

\begin{equation} \label{eq:khakhar}
\begin{split}
    & \rho \frac{\partial \left( v_{\parallel} \delta  \right)}{\partial t} + \rho v_{\parallel}^2 \frac{\partial \delta }{\partial x} = - \frac{\partial \left(\delta  \tau_{xx}\right)}{\partial x} + \tau_{yx} + \tau_{zx}\frac{\delta }{w} + \rho g \delta  \sin\theta + (\rho v_x v_{\perp})|_{y=0} \\ 
    \end{split}
\end{equation}

we get an expression for the momentum conservation in the ($x$) direction considering the interaction with the walls. Its quasi-steady state solution gives,

\begin{equation} \label{eq:momentum_final}
\begin{split}
    & \rho v_{\parallel}^2 \frac{\partial \delta }{\partial x} = - \frac{1}{2} \rho v_{\parallel}^2 \frac{\partial \delta }{\partial x} - \rho g \delta  \cos\theta\tan\theta_s - \mu_w \rho g \delta  \cos\theta \frac{\delta }{w} + \rho g \delta  \sin\theta \\ 
    & v_{\parallel} \frac{\partial \delta }{\partial x} = \frac{2 g \delta  \cos\theta \left(\tan\theta - \mu_w\frac{\delta }{w} -\tan\theta_s\right)}{3 v_{\parallel}}
\end{split}
\end{equation}
\end{widetext}

The normal stress $\tau_{xx}$ is considered as the dynamic pressure, equal to $\frac{1}{2} \rho v_{\parallel}^2 $, and its variation in the flow direction ($x$) is neglected since changes in the layer thickness are small. The shear stresses $\tau_{yx} = \rho g \delta  \cos\theta\tan\theta_s$ and $\tau_{zx} = \mu_w \rho g \delta \cos\theta$ are assumed as Coulomb's frictional stresses where $\tan\theta_s$ is the static friction of the pile $\mu_s$ and $\mu_w$ the dynamic friction coefficient of the grains with the walls. $\tau_{zx}$ is taken as suggested in \cite{Jop2005JFluidMech}. $v_x|_{y=0}$ is expected to be equal or close to $0$, so, $\rho v_x v_{\perp}$ is neglected. Combining equations (\ref{eq:mass_final}) and (\ref{eq:momentum_final}),

\begin{equation}
    L_x = \frac{3 Q_{in} \alpha \sqrt{2g h}}{2 g \bar\delta  \cos\theta \left(-\tan\theta + \mu_w\frac{\bar\delta }{w} + \tan\theta_s\right) w}
\end{equation}

where $\bar\delta = (1/L_x) \int_0^{L_x} \delta dx$. We have used the fact that $Q_{\perp} = Q_{in} = v_{\perp} L_x w $, where $L_x = \frac{X}{\cos\theta}$ is the length of the interface between the the flowing and static layers, and $v_{\parallel} = \alpha v_{in}\sin\theta = \alpha \sqrt{2g h} \sin\theta$, being $v_{in}$ the velocity of the incoming grains (free fall) and $\alpha < 1$ a dimensionless constant that accounts for the energy loss after the impact on the tip of the pile of the grains coming from the container.

At the transition ($\theta \to \theta_s$) the thickness of the fluid layer reaches a minimum $\bar\delta=\delta_{stop}$, after which the flowing grains cannot ``continually cover" the whole surface of the pile, therefore intermittency starts \cite{Lemieux2000PhysRevLett,Jop2005JFluidMech}.

\begin{equation}
    \lim_{\theta \to \theta_s} L_x = \frac{3 \alpha \sqrt{h} Q_{in}\sin\theta_s}{\sqrt{2g} \delta _{stop} \cos\theta_s \mu_w\frac{\delta_{stop}}{w} w}
\end{equation}

\begin{equation}
    X_c = \frac{3 \alpha w \sin\theta_s}{ \sqrt{2g} \mu_w \delta _{stop}^2}F_{in}h^{1/2}
\end{equation}
    
Assuming a triangular heap profile that grows with a constant angle $\theta_s$, and using the principle of mass conservation we can write the following expression for the whole heap,

\begin{equation}
F_{in} t_c = \frac{1}{2} X_c^2 \tan{\theta_s}
\label{eq.B1}
\end{equation}

Hence Eq. (\ref{eq:transition2}) can be obtained.

\bibliographystyle{unsrt}
\bibliography{submission}

\end{document}